\documentclass[iop,apj]{emulateapj}
\usepackage{graphicx,hyperref}

\shorttitle{Triangulum II}
\shortauthors{Kirby et al.}
\slugcomment{Accepted to ApJ on 2017 March 8}

\begin{document}

\newcommand{\n}{13}
\newcommand{\nm}{21}
\newcommand{\fehmeanw}{-2.24}
\newcommand{\fehmeanwerr}{0.04}
\newcommand{\fehmean}{-2.32}
\newcommand{\fehmeanerr}{0.29}
\newcommand{\fehmeanerrlower}{0.29}
\newcommand{\fehmeanerrupper}{0.29}
\newcommand{\fehsigma}{0.53}
\newcommand{\fehsigmaerr}{0.25}
\newcommand{\fehsigmaerrlower}{0.12}
\newcommand{\fehsigmaerrupper}{0.38}
\newcommand{\hiresvdiff}{2.8}
\newcommand{\hiresvdifferr}{1.7}
\newcommand{\hiresvdiffsigma}{1.7}
\newcommand{\sigmavlimone}{3.4}
\newcommand{\sigmavlimtwo}{4.2}
\newcommand{\mhalflimone}{3.7}
\newcommand{\mhalflimtwo}{5.6}
\newcommand{\mllimone}{1640}
\newcommand{\mllimtwo}{2510}
\newcommand{\vmean}{-381.7}
\newcommand{\vgsr}{-261.7}
\newcommand{\vellim}{6.1}
\newcommand{\probone}{0.47}
\newcommand{\probnm}{9.5}
\newcommand{\vrerrbig}{2.7}
\newcommand{\fehbig}{-1.40}
\newcommand{\feherrbig}{0.13}
\newcommand{\sepbigarcmin}{7.6}
\newcommand{\sepbigrh}{2.0}
\newcommand{\fehbin}{-1.91}
\newcommand{\feherrbin}{0.11}
\newcommand{\fehwbin}{-1.90}
\newcommand{\fehwerrbin}{0.09}
\newcommand{\alphafebin}{-0.48}
\newcommand{\alphafeerrbin}{0.15}
\newcommand{\alphafewbin}{-0.51}
\newcommand{\alphafewerrbin}{0.27}
\newcommand{\mgfebin}{+0.21}
\newcommand{\mgfeerrbin}{0.28}
\newcommand{\fehmr}{-1.40}
\newcommand{\feherrmr}{0.13}
\newcommand{\fehwmr}{-1.37}
\newcommand{\fehwerrmr}{0.06}
\newcommand{\alphafemr}{-0.28}
\newcommand{\alphafeerrmr}{0.23}
\newcommand{\alphafewmr}{+0.28}
\newcommand{\alphafewerrmr}{0.36}
\newcommand{\mgfemr}{+0.89}
\newcommand{\mgfeerrmr}{0.26}
\newcommand{\vright}{-375.6}
\newcommand{\vrighterr}{11.2}
\newcommand{\density}{0.0}
\newcommand{\densityerrl}{0.0}
\newcommand{\densityerru}{ 0.1}

\title{Triangulum II: Not Especially Dense After All}

\author{Evan~N.~Kirby\altaffilmark{1},
  Judith~G.~Cohen\altaffilmark{1},
  Joshua~D.~Simon\altaffilmark{2},
  Puragra~Guhathakurta\altaffilmark{3},
  Anders~O.~Thygesen\altaffilmark{1},
  Gina~E.~Duggan\altaffilmark{1}}

\altaffiltext{*}{The data presented herein were obtained at the
  W.~M.~Keck Observatory, which is operated as a scientific
  partnership among the California Institute of Technology, the
  University of California and the National Aeronautics and Space
  Administration. The Observatory was made possible by the generous
  financial support of the W.~M.~Keck Foundation.}
\altaffiltext{1}{California Institute of Technology, 1200 E.\ California Blvd., MC 249-17, Pasadena, CA 91125, USA}
\altaffiltext{2}{Observatories of the Carnegie Institution of Washington, 813 Santa Barbara Street, Pasadena, CA 91101, USA}
\altaffiltext{3}{UCO/Lick Observatory and Department of Astronomy and Astrophysics, University of California, 1156 High Street, Santa Cruz, CA 95064, USA}

\keywords{galaxies: dwarf --- Local Group --- galaxies: abundances}

%%%%%%%%%%%%%%%%%%%%%%%%%%%%%%%%%
%%%%%%%%%    ABSTRACT    %%%%%%%%
%%%%%%%%%%%%%%%%%%%%%%%%%%%%%%%%%

\begin{abstract}

%Among the Milky Way satellites discovered in the past three years, Triangulum II has presented the most difficulty in revealing its dynamical status.  Kirby et al. (2015a) identified it as the most dark matter-dominated galaxy known, with a mass-to-light ratio within the half-light radius of 3600 +3500 -2100 M_sun/L_sun. On the other hand, Martin et al. (2016) measured an outer velocity dispersion that is 3.5 +/- 2.1 times larger than the central velocity dispersion, suggesting that the system might not be in equilibrium.  From new multi-epoch Keck/DEIMOS measurements of 13 member stars in Triangulum II, we constrain the velocity dispersion to be sigma_v < 3.4 km/s (90% C.L.). Our previous measurement of sigma_v, based on six stars, was inflated by the presence of a binary star with variable radial velocity.  We find no evidence that the velocity dispersion increases with radius.  The stars display a wide range of metallicities, indicating that Triangulum II retained supernova ejecta and therefore possesses or once possessed a massive dark matter halo. However, the detection of a metallicity dispersion hinges on the membership of the two most metal-rich stars.  The stellar mass is lower than galaxies of similar mean stellar metallicity, which might indicate that Triangulum II is either a star cluster or a tidally stripped dwarf galaxy.  Detailed abundances of one star show heavily depressed neutron-capture abundances, similar to stars in most other ultra-faint dwarf galaxies but unlike stars in globular clusters.

Among the Milky Way satellites discovered in the past three years,
Triangulum~II has presented the most difficulty in revealing its
dynamical status.  \citet{kir15c} identified it as the most dark
matter-dominated galaxy known, with a mass-to-light ratio within the
half-light radius of $3600^{+3500}_{-2100}~M_{\sun}~L_{\sun}^{-1}$.
On the other hand, \citet{mar16a} measured an outer velocity
dispersion that is $3.5 \pm 2.1$ times larger than the central
velocity dispersion, suggesting that the system might not be in
equilibrium.  From new multi-epoch Keck/DEIMOS measurements of
\n\ member stars in Triangulum~II, we constrain the velocity
dispersion to be $\sigma_v < \sigmavlimone$~km~s$^{-1}$ (90\% C.L.).
Our previous measurement of $\sigma_v$, based on six stars, was
inflated by the presence of a binary star with variable radial
velocity.  We find no evidence that the velocity dispersion increases
with radius.  The stars display a wide range of metallicities,
indicating that Triangulum~II retained supernova ejecta and therefore
possesses or once possessed a massive dark matter halo.  However, the
detection of a metallicity dispersion hinges on the membership of the
two most metal-rich stars.  The stellar mass is lower than galaxies of
similar mean stellar metallicity, which might indicate that
Triangulum~II is either a star cluster or a tidally stripped dwarf
galaxy.  Detailed abundances of one star show heavily depressed
neutron-capture abundances, similar to stars in most other ultra-faint
dwarf galaxies but unlike stars in globular clusters.

\end{abstract}

%%%%%%%%%%%%%%%%%%%%%%%%%%%%%%%%%
%%%%%%%%%   SECTION 1   %%%%%%%%%
%%%%%%%%%%%%%%%%%%%%%%%%%%%%%%%%%

\section{Introduction}
\label{sec:intro}

Dwarf galaxies present excellent opportunities for studying a
multitude of aspects of galaxy formation and cosmology.  First, dwarf
galaxies are especially sensitive to stellar feedback.  They have
shallow gravitational potentials because they have little mass.  As a
result, supernovae and even winds from low-mass stars can redistribute
the metals in the galaxy's gas \citep[e.g.,][]{lar74} and even expel
metals from the galaxy \citep[e.g.,][]{dek03}.  The effect of feedback
is readily apparent in the low average metallicities of dwarf galaxies
\citep{ski89}.  Second, dwarf galaxies contain a great deal of dark
matter.  They exhibit mass-to-light ratios of tens to thousands in
solar units \citep{fab83,sim07,geh09,sim11}.  The overwhelming
dominance of dark matter relative to the luminous matter makes dwarf
galaxies the ideal targets for examining dark matter density profiles
\citep{wal07} and for searching for self-annihilation of the dark
matter particle in gamma rays \citep[e.g.,][]{drl15a}.

Measuring the dark matter content of dwarf galaxies requires precision
and delicacy.  Measuring even the total dark matter mass---not to
mention the mass profile---is nuanced.  The mass is measured by
quantifying the stellar velocity dispersion.  Smaller masses imply
shallower gravitational potentials and slower stellar orbits.
Measuring the dispersion in velocity of stars in an ultra-faint dwarf
galaxy (UFD) is difficult because it is not much larger than the
uncertainty in the measurement of a single star's radial velocity
\citep{sim07}.

Many UFDs do not have more than a few stars on the red giant branch
(RGB)\@.  Thus, the velocity dispersions must be measured from a
sample containing mostly faint stars ($V \ga 19$), precluding
high-resolution spectroscopy.  The velocity precision therefore is
limited by small samples, spectral resolution, and the random noise in
the spectrum.

Numerous systematic errors further limit the velocity precision (see
\citealt{soh07}, \citealt{sim07}, or \citealt{kir15b} for descriptions
of some of these effects).  For example, imaging spectrographs use
slits to isolate stars.  Displacement of the star from the center of
the slit will cause a shift in the measured wavelengths---and hence
velocities---relative to any reference spectrum that fills the slit,
like an arc lamp.  Another source of systematic uncertainty is the
precision of the wavelength solution.  The wavelengths can be known
only as well as the reference, usually emission lines from an arc
lamp.  The mapping from pixel position to wavelength also assumes a
model, such as a polynomial, which may be too simplistic for some
spectrographs.  Furthermore, flexure and thermal contraction of the
spectrograph during the night could cause the wavelength solution to
wander.  Finally, radial velocities are measured by comparing the
observed spectrum to a synthetic or empirical template spectrum.  A
mismatch in the shapes or strengths of absorption lines in the
template spectrum could lead to an additional source of systematic
uncertainty.

There is also one major astrophysical source of error in the velocity
dispersion.  Binary stars have orbital velocities that could exceed
the velocity dispersion of the galaxy by a factor of ten or more.
Although the center of mass of the binary system will follow galactic
orbits, the velocities of the individual binary components should not
be used in the model of the dwarf galaxy's velocity dispersion.  If a
star in a binary system is erroneously interpreted as a single star,
then its velocity may artificially inflate the measured velocity
dispersion.  Using red giants to measure the velocity dispersion is
especially prone to undetected binaries because the red giant is
probably much brighter than its companion.  The binary companion could
already have evolved past the giant phase into a white dwarf, or the
companion might still be on the main sequence.  For a 12~Gyr
population, like an ancient dwarf galaxy, only stars with masses
within $\sim 1\%$ of each other will be on the RGB simultaneously
\citep[e.g.,][]{dem04}.  For other binary systems, the red giant is
likely to be many times brighter than its companion.  Therefore, the
only way to determine the binarity of most red giants in dwarf
galaxies is multi-epoch spectroscopy to search for variability in
radial velocity.

The binary frequency in dwarf galaxies is not well known.
\citet{min13} estimated the binary fraction of four classical dwarf
spheroidal galaxies (dSphs) using spectroscopic data to be
$46^{+13}_{-9}\%$, consistent with the Milky Way (MW) field
population.  However, he also found that the fraction could vary among
dSphs.  For example, the binary fraction in the Carina dSph
($14^{+28}_{-5}\%$) was found to be inconsistent with the MW field
population at 90\% confidence.  M.E. Spencer et al.\ (in prep.) found
a binary fraction of approximately $(30 \pm 10)\%$ in the Leo~II dSph.
They estimated that such a binary fraction in an UFD with a velocity
dispersion of 2.0~km~s$^{-1}$ could spuriously inflate the velocity
dispersion on average to 2.7~km~s$^{-1}$, leading to an 80\%
overestimate in the galaxy's mass.  Unfortunately, the binarity of the
stellar populations in UFDs is even less well measured than classical
dSphs.  From \textit{Hubble Space Telescope} photometry, \citet{geh13}
measured a binary fraction of ($47^{+16}_{-14}\%$) in the Hercules UFD
and ($47^{+37}_{-17}\%$) in the Leo~IV UFD\@.  Only one UFD star (in
Hercules) has a complete binary orbit \citep{koc14a}.  Other binaries
are known in Bo{\"o}tes~I \citep{kop11} and Bo{\"o}tes~II
\citep{ji16_boo}.

In this article, we critically examine the previously published
spectroscopic properties of the Triangulum~II UFD (Tri~II), paying
particular attention to the dispersion and spatial distribution of
stellar radial velocities.  \citet{lae15a} discovered Tri~II at a
distance of $30 \pm 2$~kpc in the Panoramic Survey Telescope and Rapid
Response System \citep[Pan-STARRS;][]{kai10} photometric survey.  Its
luminosity is just $450~L_{\sun}$, and its 2D half-light radius is
only 34~pc.  \citet[][hereafter K15a]{kir15c} measured its velocity
dispersion from medium-resolution, multi-object spectroscopy to be
$\sigma_v = 5.1^{+4.0}_{-1.4}$~km~s$^{-1}$ from six stars near the
galaxy's center.  \citet[][hereafter M16]{mar16a} corroborated this
central velocity dispersion and found that it increased to as much as
14~km~s$^{-1}$ in the outermost parts of the galaxy.  Such a galaxy
would almost certainly be out of dynamical equilibrium.  If Tri~II is
in dynamical equilibrium, either measurement of the velocity
dispersion would make it the most dark matter-dominated galaxy known.
As a result, it would be an excellent candidate for the indirect
detection of dark matter from self-annihilation \citep{gen16}.

The chemical properties of Tri~II suggest a very metal-poor stellar
system ($\langle {\rm [Fe/H]} \rangle \approx -2.5$).  K15a found that
the metallicities of three stars are inconsistent with being
identical.  As a result, Tri~II chemically enriched itself, a hallmark
of a bona fide galaxy rather than a star cluster \citep{wil12}.
\citet[][hereafter V17]{ven17} additionally obtained high-resolution
spectra of two stars in Tri~II\@.  The two stars differ in iron
abundance by ($0.4 \pm 0.3$)~dex.  The magnesium and calcium abundance
are discrepant at $\sim 2\sigma$.  Thus, these two high-resolution
spectra support K15a's claim that Tri~II is chemically diverse.

We describe our observations in Section~\ref{sec:obs}.
Sections~\ref{sec:rv} and \ref{sec:metallicity} detail our
medium-resolution spectroscopic measurements of velocities and
metallicities, respectively.  We additionally obtained a
high-resolution spectrum of one star, as described in
Section~\ref{sec:detailed}.  Section~\ref{sec:discussion} closes with
a discussion of our findings.

%%%%%%%%%%%%%%%%%%%%%%%%%%%%%%%%%
%%%%%%%%%   SECTION 2   %%%%%%%%%
%%%%%%%%%%%%%%%%%%%%%%%%%%%%%%%%%

\section{Observations}
\label{sec:obs}

\subsection{Keck/DEIMOS}
\label{sec:deimos}

\begin{deluxetable*}{lcccccc}
  \tablewidth{0pt}
  \tablecolumns{7}
  \tablecaption{DEIMOS Observations\label{tab:obs}}
  \tablehead{\colhead{Slitmask} & \colhead{Targets} & \colhead{UT Date} & \colhead{Airmass} & \colhead{Seeing} & \colhead{Exposures} & \colhead{Exp.\ Time} \\
    \colhead{ } & \colhead{ } & \colhead{ } & \colhead{ } & \colhead{($''$)} & \colhead{ } & \colhead{(s)}}
  \startdata
  TriIIb\tablenotemark{a} & 29 & 2015 Oct 8\phn & 1.4 & 0.9 & 3 & 3120 \\
  TriIIc                  & 25 & 2015 Dec 14    & 1.1 & 0.9 & 4 & 7200 \\
  TriIId                  & 28 & 2016 Jan 29    & 1.1 & 0.9 & 4 & 5100 \\
  TriIIe                  & 25 & 2016 Jan 29    & 1.3 & 1.1 & 3 & 4803 \\
  TriIIf                  & 21 & 2016 Jun 29    & 1.6 & 1.0 & 2 & 2700 \\
  TriIIg                  & 27 & 2016 Sep 7\phn & 1.1 & 0.7 & 3 & 3060 \\
  \enddata
  \tablenotetext{a}{Observations by K15a.}
\end{deluxetable*}

The Deep Extragalactic Imaging Multi-Object Spectrograph
\citep[DEIMOS;][]{fab03} is a medium-resolution spectrograph at the
Nasmyth focus of the Keck~II telescope.  The multi-object mode employs
custom-milled aluminum slitmasks designed for a specific pointing and
position angle.  Each target has a unique slit cut for it.  Due to the
limited target density, a single slitmask could not accommodate all
the stars identified as members by both K15a and M16.  Therefore, the
target lists for the slitmasks were not identical.  We designed six
different slitmasks observed on five different dates (see
Table~\ref{tab:obs}).  The targets were selected from the union of the
member lists of K15a and M16.  Some additional stars known to be
non-members were also targeted.

The spectrographic configuration was the same for all six slitmasks.
We used the 1200G grating, which has a groove spacing of
1200~mm$^{-1}$ and a blaze wavelength of 7760~\AA\@.  This grating
provided a spectral resolving power of $R = 7000$ at 8500~\AA\@.  The
central wavelength was set to be 7800~\AA, yielding a typical
wavelength range of 6500~\AA\ to 9000\AA, but the range varied from
star to star depending on the exact location of the slit on the
slitmask.  We used the OG550 order-blocking filter to isolate the
light diffracted from the first order of the grating.  We used a
quartz lamp for flat fielding and Ne, Ar, Kr, and Xe arc lamps for
wavelength calibration.  DEIMOS's flexure compensation system provided
extra wavelength stability.

We reduced the DEIMOS spectra using \texttt{spec2d}
\citep{coo12,new13}, a data reduction pipeline built by the Deep
Extragalactic Exploration Probe (DEEP) team.  We used our own
modifications to the pipeline \citep{kir15b,kir15a}.  These
modifications included an improvement to the way in which sky lines
are traced along the slit, and the extraction was improved by taking
into account differential atmospheric refraction.  The final products
of the pipeline are flat-fielded, wavelength-calibrated,
sky-subtracted spectra and their associated variance arrays.

\begin{figure*}[t!]
  \centering
  \includegraphics[width=0.8\textwidth]{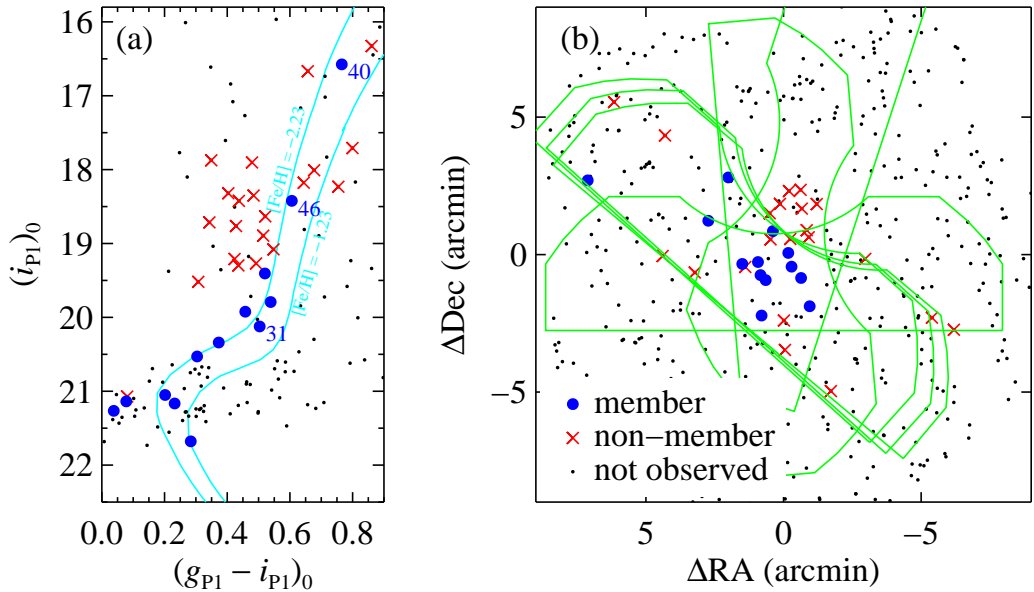}
  \caption{({\textit a}) CMD from Pan-STARRS photometry.
    Spectroscopic members are shown as large blue points, and
    non-members are red crosses.  The cyan curves show 14~Gyr Padova
    isochrones \protect \citep{gir04} at the distance of Tri~II\@.
    The two isochrones have metallicities of ${\rm [Fe/H]} = -1.23$
    and $-2.23$.  ({\textit b}) The map of spectroscopic targets.  The
    DEIMOS slitmask outlines are shown in green.\label{fig:cmd_map}}
\end{figure*}

We adopted photometry from the first data release of Pan-STARRS
\citep{cha16,fle16}.  Figure~\ref{fig:cmd_map} shows the observed
stars in the CMD and celestial coordinates.  For the purposes of
estimating effective temperatures and surface gravities
(Section~\ref{sec:metallicity}), we converted Pan-STARRS magnitudes to
the Sloan Digital Sky Survey (SDSS) system following \citet{ton12}.
The figures and tables in this article present the un-transformed,
native Pan-STARRS photometry.  We corrected magnitudes for reddening
and extinction using the dust maps of \citet{sch98}.

In this article, we refer to the stars by the identification given by
M16 except for star~128, which was not in M16's sample.

\subsection{Keck/HIRES}
\label{sec:hires}

We obtained Keck/HIRES spectra of star 40, the brightest known member
of Tri~II, on 2016 Aug 25.  Figure~\ref{fig:cmd_map}a identifies star
40 in the CMD\@.  This star is also the subject of the high-resolution
spectroscopic analysis by V17.  We obtained 7 exposures of 1800~s each
for a total exposure time of 3.5~hr.  The seeing was $0.5''$.  We used
the $7.0'' \times 0.86''$ slit, which provided a spectral resolving
power of $R = 49000$ over the spectral range 3933~\AA\ to
8374~\AA\ with some inter-order gaps redward of 6410~\AA\@.

We used a quartz lamp for flat fielding and a ThAr arc lamp for
wavelength calibration.  We reduced the raw frames into 1D spectra
using version 5.2.4 of the MAKEE data reduction
pipeline\footnote{\url{http://www.astro.caltech.edu/~tb/makee/}}.
This pipeline produces a flat-fielded, wavelength-calibrated 1D
spectrum and associated variance array for each exposure.  We summed
the seven individual spectra to produce a single spectrum.  We also
constructed the error spectrum by summing the individual error spectra
in quadrature.

The HIRES spectrum is composed of 47 observed echelle orders, some of
which overlap in wavelength.  We combined the orders into one
continuous spectrum except for the inter-order gaps in the redder part
of the spectrum.  Before we could combine them, we needed to remove
the response function, which caused the edges of orders to have fewer
counts than the middle of orders.  We fit a separate spline to each
order.  The splines had breakpoint spacings of 500 pixels.  After
dividing by the spline, the order appeared flat.  We then interpolated
all the orders onto a uniform wavelength array.  We averaged
overlapping parts of different orders with inverse variance weighting
to maximize S/N\@.  We calculated the final spectrum's S/N as the
standard deviation of the continuum-divided spectrum in the relatively
line-free spectral region from 5720~\AA\ to 5800~\AA\@.  The result is
$\rm{S/N} = 53$~pixel$^{-1}$ or 103 per spectral resolution element.

%%%%%%%%%%%%%%%%%%%%%%%%%%%%%%%%%
%%%%%%%%%   SECTION 3   %%%%%%%%%
%%%%%%%%%%%%%%%%%%%%%%%%%%%%%%%%%

\section{Kinematics}
\label{sec:rv}

\addtocounter{table}{1}

We measured radial velocities from the DEIMOS spectra in the same
manner as K15a and \citet{kir15b}.  We refer the reader to those
articles for more detail on the technique.  Briefly, we matched the
observed spectra to template spectra.  We varied the radial velocity
of the template spectra to minimize $\chi^2$.  We also correcting for
mis-centering in the slit by shifting the spectrum to ensure that
telluric features, such as the Fraunhofer A and B bands, were at zero
velocity.  We estimated velocity errors by re-measuring the radial
velocity with the best fitting template but with noise added to the
observed spectrum.  We took the random velocity uncertainty to be the
standard deviation among 1000 noise trials.  As found previously
\citep{sim07,kir15b}, the random noise is an incomplete description of
the velocity error.  We added a systematic error component of
1.49~km~s$^{-1}$ in quadrature with the random error to estimate the
total error.  Table~\ref{tab:velocity} gives the coordinates,
photometry, velocities, and membership of each star.  The $VI$
photometry is from the Keck/LRIS observations by K15a, and the $gi$
photometry is from the Pan-STARRS observations by M16.  For stars
observed multiple times, the velocity quoted is the mean of the
individual observations weighted by the inverse variance.  The
velocities are in the heliocentric frame.

\subsection{Membership}
\label{sec:membership}

We used roughly the same membership criteria as K15a.  K15a used a
velocity cut based on the velocity dispersion.  However, we show in
Section~\ref{sec:vdisp} that we could not resolve the velocity
dispersion of Tri~II\@.  Therefore, we accept as members any stars
within 50~km~s$^{-1}$ of the mean velocity that also pass the other
membership cuts.  We also checked that no member star had a strong
\ion{Na}{1}~$\lambda 8190$ doublet, which would indicate that the star
is a foreground dwarf.  These criteria did not rule out any star
previously classified as a member by K15a and M16.  The data set
contains 13 stars identified as members by K15a and M16: 1 from K15a,
7 from M16, and 5 from both.

Additionally, we found that star~26, with a velocity of $\vright \pm
\vrighterr$~km~s$^{-1}$, is a member.  M16 measured $-84.1 \pm
8.2$~km~s$^{-1}$ and therefore classified it as a non-member.  Star~8,
classified as a member by M16, is not in our sample.  Star~25 is a
double-lined spectroscopic binary.  Its DEIMOS spectrum has a
bifurcated H$\alpha$ absorption profile in exposures from both the
TriIIc and TriIId slitmasks.  The velocities of the individual
H$\alpha$ lines straddle the mean velocity of Tri~II\@.  Although it
is probably a member of Tri~II, we did not use it in the determination
of the velocity dispersion because our incomplete temporal sampling
did not allow us to determine the center-of-mass velocity reliably.
The resulting sample size of candidate members (excluding star~25) is
\n\@.

Star~31 deserves some extra scrutiny.  It has the highest metallicity
(${\rm [Fe/H]} = \fehbig \pm \feherrbig$; see
Section~\ref{sec:metallicity}) and largest separation from the center
of the galaxy ($r = \sepbigarcmin' = \sepbigrh r_h$) of any of the
member stars.  It is also the only star whose $1\sigma$ velocity error
bar does not overlap the mean velocity of Tri~II, which is consistent
with expectations for a sample of 13 stars with Gaussian measurement
uncertainties.  We estimated the probability that star~31 is a
non-member by querying the Besan{\c c}on model of Galactic structure
and kinematics \citep{rob03}.  We selected only stars at the Galactic
coordinates of Tri~II with colors and magnitudes within 0.03~mag of
star~31.  In a large sample of synthetic stars, $\probone\%$ had
radial velocities within $\vellim$~km~s$^{-1}$ of Tri~II's mean
velocity, where $\vellim$~km~s$^{-1}$ is the sum of the 90\%
C.L. upper limit on $\sigma_v$ (calculated below) and the velocity
uncertainty on star~31.  In our sample of \nm\ stars classified as
non-members by our DEIMOS spectroscopy, the probability of finding a
non-member that falls in the velocity window is $\mathbf{1 - (1 -
  0.0047)^{\nm} = \probnm\%}$.  The CMD position of star~31
(Figure~\ref{fig:cmd_map}a) is consistent with its measured
metallicity.  The star also has a radial velocity and
\ion{Na}{1}~$\lambda 8190$ line strength consistent with a red giant
member of Tri~II\@.  Therefore, this star passes all of our membership
criteria, and we retain it in our sample.  We discuss its influence on
the metallicity dispersion in Section~\ref{sec:metallicity}.

\subsection{Validation and Radial Trends}

We compared our DEIMOS velocities with two other sources: our HIRES
spectrum and M16.  The HIRES velocity of star~40 is $-378.8 \pm
0.2$~km~s$^{-1}$, which is $\hiresvdiff \pm
\hiresvdifferr$~km~s$^{-1}$ higher than the DEIMOS velocity.  There
are several possible causes for the discrepancy.  First, there is a
9\% chance that the difference arises from random noise, which is
almost entirely dominated by the DEIMOS spectrum.  Second, there could
be a zeropoint offset in the wavelength calibration of HIRES versus
DEIMOS\@.  Third, we might have underestimated the error in the DEIMOS
velocity measurement.  Regardless, the discrepancy is not very large,
and offsets of this magnitude do not affect our conclusions.

\begin{figure}[t!]
  \centering
  \includegraphics[width=\columnwidth]{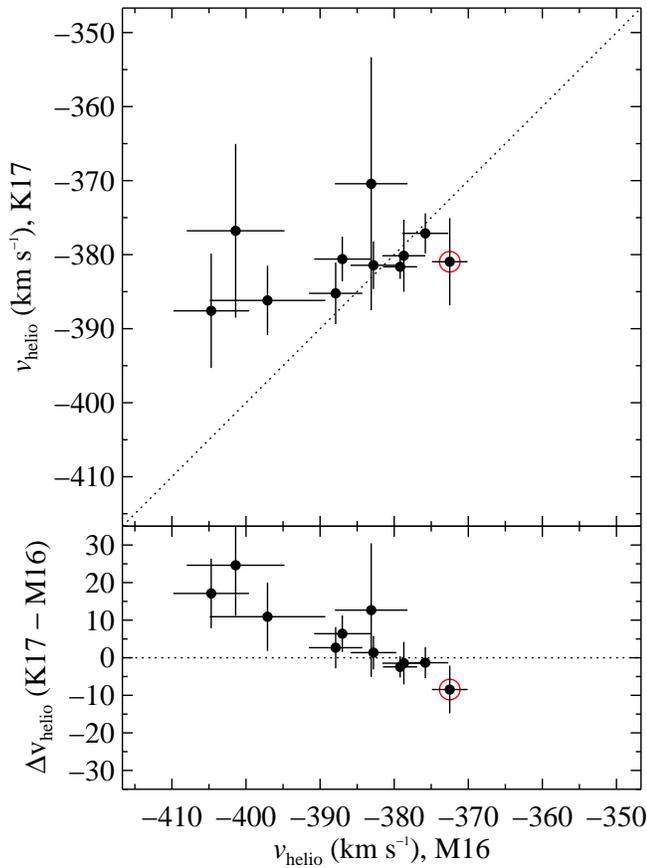}
  \caption{The velocity measurements from this work (K17) compared to
    M16.  The dotted lines show where the two measurements would be
    equal.  Star~46, the binary, is indicated by a red circle.  The
    K17 value of the velocity for the star is the mean of our four
    measurements.  Star~26 is not shown because M16 measured $v_{\rm
      helio} = -84.1 \pm 8.2$~km~s$^{-1}$, whereas we measured $v_{\rm
      helio} = \vright \pm
    \vrighterr$~km~s$^{-1}$.\label{fig:comparev}}
\end{figure}

Our sample overlaps with that of M16 by 27 stars.  We count 12 of
these stars as members, excluding star~25, the double-lined
spectroscopic binary.  M16 did not consider one of these, star~26, a
member because they measured a much different radial velocity than we
did.  Figure~\ref{fig:comparev} shows our measurements of radial
velocity for the remaining 11 stars compared to M16.  We determined
star~46 to exhibit a radial velocity that varies in time
(Section~\ref{sec:velvar}).  Among the remaining sample of ten stars,
our measurements of radial velocity for six stars overlap with M16's
measurements within the quadrature sum of the $1\sigma$ error bars.
For perfectly estimated errors, we would have expected 68\% of the
sample, or about seven stars, to overlap.  Another way to compare
consistency between the samples is to examine the quantity
$(v_{\rm{K17}}-v_{\rm{M16}})/\sqrt{\delta v_{\rm{K17}}^2+\delta
  v_{\rm{M16}}^2}$.  For perfect errors, this quantity would be
normally distributed with a mean of $(0 \pm 1)$ and a standard
deviation of $(1 \pm 0.18)$, assuming $N=10$.  We measured a mean of
$-0.63$ and a standard deviation of $0.93$.  Thus, the samples are
roughly consistent after excluding stars~25, 26, and 46.  However, as
we show in Section~\ref{sec:vdisp}, the consistency between the two
samples does not guarantee that we infer the same properties of the
underlying velocity distribution.

\begin{figure}[t!]
  \centering
  \includegraphics[width=\columnwidth]{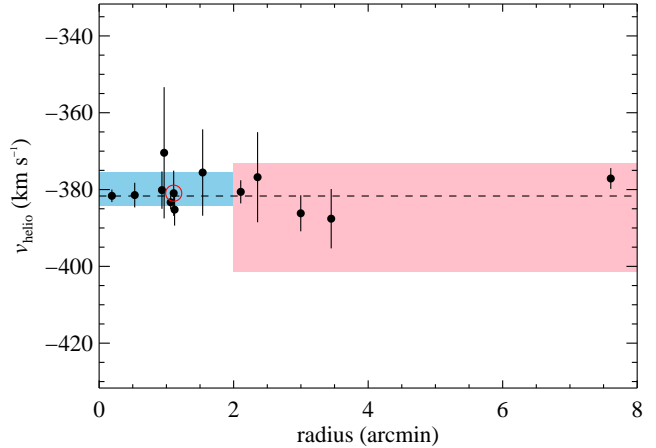}
  \caption{Radial velocity of Tri~II members versus their distances
    from the center of the galaxy.  The dashed line shows the mean
    velocity.  The shaded regions show the velocity dispersions in
    different regions of the galaxy published by M16.  The mean
    velocity of star~46, the binary, is indicated by a red circle.
    Our measurements do not show evidence for a velocity dispersion
    changing with radius.\label{fig:vdist}}
\end{figure}

Our updated velocities affect M16's conclusion that the velocity
dispersion increases with radius.  Figure~\ref{fig:vdist} shows our
measurements of velocity versus angular distance from the center of
Tri~II\@.  M16 claimed that the velocity dispersion increased from
$4.4^{+2.8}_{-2.0}$~km~s$^{-1}$ in the central $2\arcmin$ (blue shaded
region) to $14.1^{+5.8}_{-4.2}$~km~s$^{-1}$ outside of $2\arcmin$ (red
shaded region).  There is no apparent increase of velocity dispersion
or change in mean velocity with radius in our data.  In fact, the
velocity of every member star in our sample except star~31 is
consistent with the mean velocity within $1\sigma$.  We note that our
exclusion of star~25, the double-lined spectroscopic binary, is partly
responsible for our finding of a smaller velocity dispersion at $r >
2\arcmin$ than M16 found.  However, our revisions to the velocities of
the three stars mentioned in the previous paragraph are the primary
cause for our downward revision of the outer velocity dispersion.

\subsection{Velocity Variability}
\label{sec:velvar}

\begin{figure}[t!]
  \centering
  \includegraphics[width=\columnwidth]{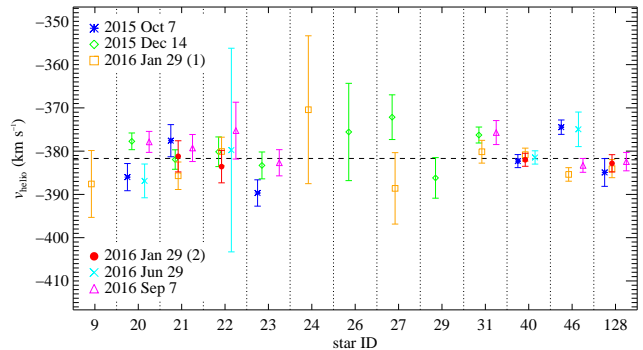}
  \caption{Heliocentric radial velocities observed at six different
    epochs.  Each vertical column represents the measurements for one
    star.  Within the column, the stars are ordered according to the
    date of observation.  There are two independent measurements on
    2016 Jan 26.\label{fig:var}}
\end{figure}

The advantage of observing stars over multiple epochs is that their
radial velocities can be monitored.  Any change in velocity in excess
of the measurement uncertainty likely indicates binarity.
Figure~\ref{fig:var} shows the individual velocity measurements of the
12 member stars.  Each column is the velocity curve of a unique member
star.  The horizontal spacing between measurements within a column is
proportional to the time elapsed between observations.

\begin{figure}[t!]
  \centering
  \includegraphics[width=\columnwidth]{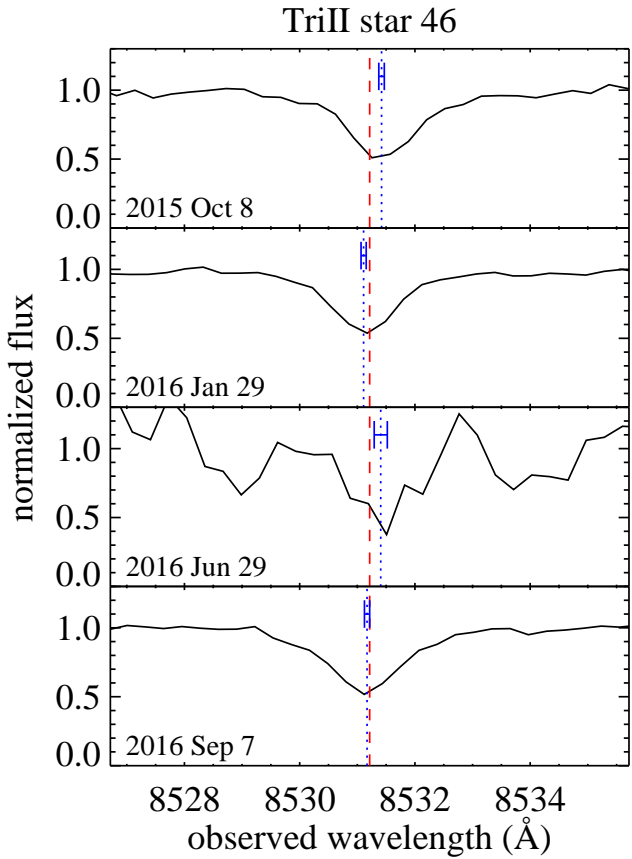}
  \caption{Four observations of star~46.  Only a small region of the
    spectrum around \ion{Ca}{2}~$\lambda$8542 is shown.  The
    wavelengths are corrected for slit mis-centering, and they are
    shown in the heliocentric frame.  The red dashed line shows the
    expected center of the line at the mean velocity of Tri~II:
    $\langle v_{\rm helio} \rangle = \vmean$~km~s$^{-1}$.  The blue
    dotted lines show the center of the line at the measured velocity
    for each observation.  The blue whiskers surrounding the blue
    dotted lines reflect the uncertainty in the position of the line
    center.  The four observations show significant variation,
    indicating that the star's velocity is
    variable.\label{fig:spectra}}
\end{figure}

We quantified the velocity variability for each star as $\sigma(v) =
{\rm stddev}(v_i/\delta v_i)$.  Only star~46 had $\sigma(v) > 2$.
Figure~\ref{fig:spectra} demonstrates the shift in the observed
wavelength of \ion{Ca}{2}~$\lambda 8542$ in star~46 for four different
dates.  The changing wavelength is apparent even by eye.  V17 also
identified star~46 as a velocity variable from high-resolution
spectroscopy.  However, their measurement of the velocity amplitude
($24.5 \pm 2.1$~km~s$^{-1}$) is significantly higher than the
10.9~km~s$^{-1}$ peak-to-peak amplitude that we observed with
DEIMOS\@.  Our temporal sampling is not dense enough to identify the
true radial velocity amplitude with any certainty.  V17's third of
three velocity measurements for star~46 occurred between the dates of
our first and second measurements.  It is possible that the star's
orbit is such that V17 measured the velocity close to the peak,
whereas our measurements happened to be spaced closer to the midpoint
of the velocity curve.

Our four DEIMOS radial velocity samples of star~46 are consistent with
a period of $\sim 220$~days and a semi-amplitude of $\sim
5$~km~s$^{-1}$.  Assuming an edge-on, circular orbit and a primary
mass of $0.8~M_{\sun}$, the mass of the secondary and the separation
are uniquely constrained to be $0.14~M_{\sun}$ and 0.7~AU\@.
Decreasing the inclination ($i < 90$) would increase the secondary
mass and slightly increase the separation.  If the semi-amplitude is
instead 12~km~s$^{-1}$, as indicated by V17, then the secondary mass
and separation would be $0.38~M_{\sun}$ and 0.8~AU for an edge-on
system.  These ranges of mass ratios and separations are well within
the known ranges for stellar multiplicity in the solar neighborhood
\citep{duc13}.

\subsection{Velocity Dispersion}
\label{sec:vdisp}

Because K15a measured radial velocities during only one epoch, we did
not know that star~46---one of only six member stars in that
article---was variable in velocity.  K15a pointed out that removing
this star (called star~65 in that article) resulted in a velocity
dispersion of $\sigma_v = 2.8^{+4.0}_{-1.7}$~km~s$^{-1}$ based on five
stars.  This is a marginally resolved velocity dispersion.  Thus, the
binary orbital velocity of star~46 was driving the measurement of the
velocity dispersion of Tri~II rather than orbits in the galaxy's
potential.  In other words, star~46 was responsible for K15a's
erroneously high value of $\sigma_v$.

Here, we revise our estimate of $\sigma_v$, taking into account the
newly discovered binarity of star~46 and the expanded sample of member
stars with radial velocities.  We re-measured velocity dispersion in
the same manner as K15a, excluding star~46.  For stars with multiple
measurements, we took the average of $v_{\rm helio}$ weighted by the
inverse variance of the individual measurements.  Our method of
measuring $\sigma_v$ is based on \citet{wal06} and is described fully
by K15a and \citet{kir15b}.  Briefly, we define a likelihood function
\citep[Equation~1 of][]{kir15b} that the velocity measurements and
their errors, $(v_{\rm helio})_i$ and $(\delta v_{\rm helio})_i$, are
described by a mean velocity, $\langle v_{\rm helio} \rangle$, and a
dispersion, $\sigma_v$.  We maximized the likelihood with a Monte
Carlo Markov Chain (MCMC) with $10^7$ links.  We used a
Metropolis--Hastings algorithm to perform the optimization.

\begin{figure}[t!]
  \centering
  \includegraphics[width=0.9\columnwidth]{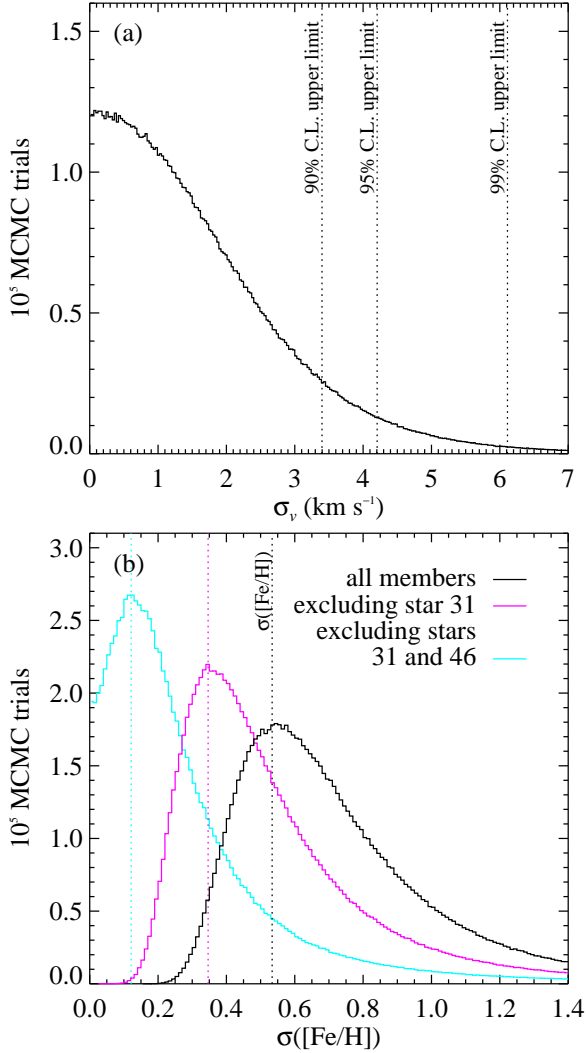}
  \caption{The posterior probability distribution on ({\textit a}) the
    velocity dispersion and ({\textit b}) metallicity dispersion.  The
    histograms show the number of accepted MCMC trials in bins of
    0.03~km~s$^{-1}$ for $\sigma_v$ and 0.01~dex for $\sigma({\rm
      [Fe/H]})$.  The vertical dashed lines indicate confidence levels
    for the upper limit on $\sigma_v$ in the top panel and the mean
    value for $\sigma({\rm [Fe/H]})$ in the bottom panel.  The
    metallicity dispersion is computed for all candidate member stars
    (\textit{black}), excluding star~46 (\textit{magenta}), and
    excluding both star~46 and star~31 (\textit{cyan}).  Star~46 is a
    likely member, but it is a binary and very $\alpha$-poor.  Star~31
    is far from the galaxy center, and it is the most metal-rich star
    in the sample, but it passes every membership cut
    (Section~\ref{sec:membership}).\label{fig:sigmav}}
\end{figure}

\begin{deluxetable}{lr@{ }l}
\tablewidth{0pt}
\tablecolumns{3}
\tablecaption{Properties of Triangulum~II\label{tab:properties}}
\tablehead{\colhead{Property} & \multicolumn{2}{c}{Value}}
\startdata
$N_{\rm member}$ & 13 & \\
$\log (L_{V}/L_{\sun})$ & $2.65 \pm 0.20$ & \\
$r_h$ & $3.9_{-0.9}^{+1.1}$ & arcmin \\
$r_h$ & $ 34_{-  8}^{+  9}$ & pc \\
$\langle v_{\rm helio} \rangle$ & $-381.7 \pm 1.1$ & km~s$^{-1}$ \\
$v_{\rm GSR}$ & $-261.7$ & km~s$^{-1}$ \\
$\sigma_v$ & $<3.4$ & km~s$^{-1}$ (90\% C.L.) \\
           & $<4.2$ & km~s$^{-1}$ (95\% C.L.) \\
$M_{1/2}$ & $<3.7 \times 10^5$ & km~s$^{-1}$ (90\% C.L.) \\
          & $<5.6 \times 10^5$ & km~s$^{-1}$ (95\% C.L.) \\
$(M/L_V)_{1/2}$\tablenotemark{a} & $<1640$ & $M_{\sun}~L_{\sun}^{-1}$ (90\% C.L.) \\
                                 & $<2510$ & $M_{\sun}~L_{\sun}^{-1}$ (95\% C.L.) \\
$\rho_{1/2}$\tablenotemark{b} & $<2.2 $ & $M_{\sun}~{\rm pc}^{-3}$ (90\% C.L.) \\
                              & $<3.3 $ & $M_{\sun}~{\rm pc}^{-3}$ (95\% C.L.) \\
$\langle {\rm [Fe/H]} \rangle$ & $-2.24 \pm 0.05$ & \\
\enddata
\tablenotetext{a}{Mass-to-light ratio within the half-light radius, calculated as $M_{1/2} = 4G^{-1}\sigma_v^2 r_h$ \citep{wol10}.}
\tablenotetext{b}{Density within the half-light radius.}
\tablerefs{The measurements of $\log L_V$ and $r_h$ come from \citet{lae15a}.}
\end{deluxetable}
\addtocounter{table}{1}

The top panel of Figure~\ref{fig:sigmav} shows the distribution of
$\sigma_v$ in the $10^7$ MCMC trials.  The distribution rises all the
way to zero velocity dispersion.  Thus, we could not resolve a
velocity dispersion of Tri~II\@.  Instead, we estimated its upper
limit.  The 90\% C.L.\ upper limit, enclosing 90\% of the MCMC trials,
is $\sigma_v < \sigmavlimone$~km~s$^{-1}$.  The 95\% C.L.\ upper limit
is $\sigma_v < \sigmavlimtwo$~km~s$^{-1}$.  These values are also
reported in Table~\ref{tab:properties}.  Tri~II joins Segue~2,
Bo{\"o}tes~II, and Tucana~III in the group of UFDs with spectroscopy
of multiple stars but without a detection of the velocity dispersion
\citep{kir13a,ji16a,sim16}.  These upper limits are consistent with
K15a's value of $\sigma_v$ when star~46 is removed from the sample.
The MCMC distribution of $\sigma_v$ from that sample of just five
stars is not significantly distinct from zero.  Thus, neither we nor
K15a can resolve the velocity dispersion of Tri~II\@.

The mass within the half-light radius of a spheroidal galaxy is
proportional to the square of the velocity dispersion.  Therefore, the
downward revision to $\sigma_v$ implies a downward revision to the
mass within the half-light radius of Tri~II, which K15a estimated to
be $M_{1/2} = (8.9^{+6.8}_{-3.0}) \times 10^5~M_{\sun}$.  In fact, we
can estimate only an upper limit to the mass.  Using
\citeauthor{wol10}'s (\citeyear{wol10}) formula for dynamical mass, we
constrain the mass within the 3D half-light radius to be $M_{1/2} <
\mhalflimone \times 10^5~M_{\sun}$ (90\% C.L.)  or $M_{1/2} <
\mhalflimtwo \times 10^5~M_{\sun}$ (95\% C.L.).

Our sample overlaps M16's sample by ten member stars after excluding
stars~25, 26, and 46.  We repeated the estimation of $\sigma_v$ on
M16's measurements of velocities for these ten stars.  We resolved the
velocity dispersion as $\sigma_v = 9.1^{+3.9}_{-2.0}$~km~s$^{-1}$,
consistent with M16's estimation.  Therefore, the different result
obtained from our sample versus M16's sample is not strictly a result
of the exclusion of certain stars.  Instead, our higher-S/N velocity
measurements themselves are different enough from M16 to reduce
$\sigma_v$ to an undetectable level.

\begin{figure}[t!]
\centering
\includegraphics[width=\columnwidth]{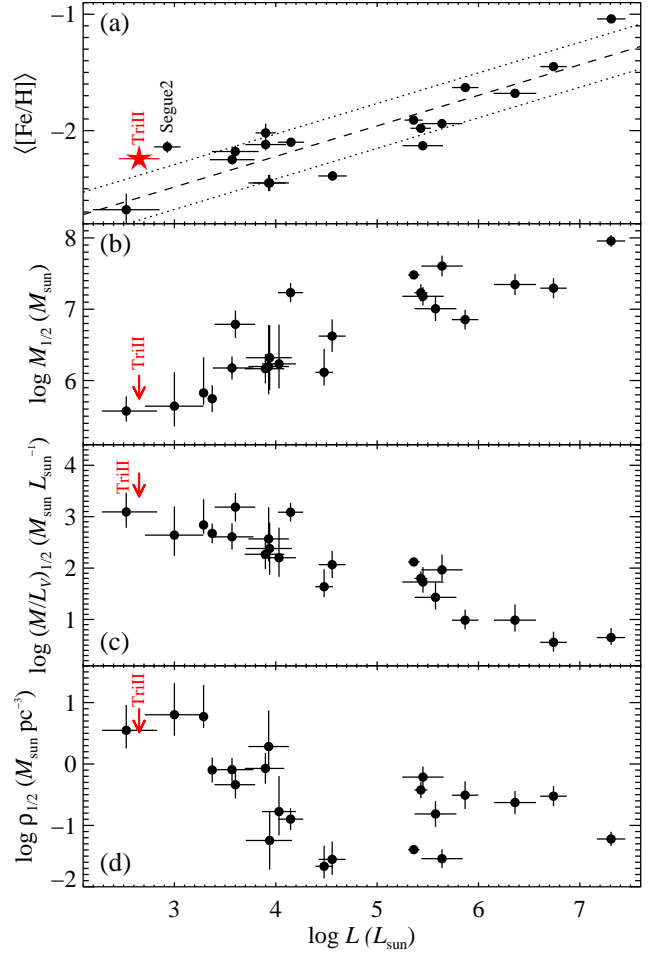}
\caption{({\textit a}) Luminosity--metallicity relation.  The dashed
  line shows the linear fit to the galaxies except Tri~II
  \citep{kir13b,kir15b}, and the dotted lines show the rms dispersion
  about the fit.  ({\textit b}) Masses of MW satellite galaxies within
  their 3D half-light radii assuming dynamical equilibrium.  Data are
  from \citet[][and references therein]{mcc12}, \citet{sim15},
  \citet{kop15b}, and \citet{kir15b}.  ({\textit c}) Mass-to light
  ratios within the 3D half-light radii.  ({\textit d}) Mass densities
  within the 3D half-light radii.\label{fig:trends}}
\end{figure}

This new mass limit indicates that Tri~II does not have the largest
mass-to-light ratio ($M/L$) of any non-disrupting galaxy.  Instead,
the upper limit on $M/L$ is consistent with the envelope of $M/L$
versus luminosity defined by other dwarf galaxies
(Figure~\ref{fig:trends}c).  Furthermore, the amount of dark matter in
Tri~II is less than previously estimated.  As a result, Tri~II is not
likely to be a significantly better candidate for indirect dark matter
detection than other UFDs of similar luminosity.

%%%%%%%%%%%%%%%%%%%%%%%%%%%%%%%%%
%%%%%%%%%   SECTION 4   %%%%%%%%%
%%%%%%%%%%%%%%%%%%%%%%%%%%%%%%%%%

\section{Metallicity and [$\alpha$/Fe]}
\label{sec:metallicity}

In addition to radial velocities, we also measured metallicities and
detailed abundances from the DEIMOS spectra.  To maximize S/N for
abundance measurements, we coadded the 1D spectra of the nine member
stars that had multiple observations.  For star~46, we removed the
Doppler shift due to the binary orbital velocity before coaddition.
We measured atmospheric parameters for the stars using spectral
synthesis, as described by \citet{kir08a,kir10}.  We used the known
distance, $30 \pm 2$~kpc \citep{lae15a} in combination with Padova
isochrones \citep{gir04} to determine the stars' surface gravities and
a first guess at their effective temperatures ($T_{\rm eff}$).  We
measured $T_{\rm eff}$, [Fe/H], [Mg/Fe], [Si/Fe], [Ca/Fe], [Ti/Fe],
and [$\alpha$/Fe] by fitting the observed, continuum-divided spectra
to a grid of synthetic spectra.  The ratio [$\alpha$/Fe] is not an
arithmetic average of the ratios of individual $\alpha$ elements.
Instead, it is measured by minimizing the $\chi^2$ of the fit to all
$\alpha$ element absorption lines simultaneously.
Table~\ref{tab:metallicity} lists the results for the six stars for
which it was possible to measure abundances.  The solar reference
scale is from \citet{asp09}\footnote{$12 + \log n({\rm Mg})/n({\rm H})
  \equiv \epsilon({\rm Mg}) = 7.58$, $\epsilon({\rm Si}) = 7.55$,
  $\epsilon({\rm Ca}) = 6.36$, $\epsilon({\rm Ti}) = 4.99$, and
  $\epsilon({\rm Fe}) = 7.50$}.  The other member stars were too faint
or their lines were too weak due to their warm temperatures.
Table~\ref{tab:properties} gives the mean of the stellar metallicities
weighted by their inverse variances.

We also measured metallicities from the near-infrared Ca triplet (CaT)
following \citet{sta10}.  This method uses the strength of
\ion{Ca}{2}~$\lambda 8498$ and \ion{Ca}{2}~$\lambda 8542$ coupled
with the absolute magnitude of the star (as a proxy for the
temperature and gravity of the star) to estimate the star's
metallicity.  The calibration is based on iron abundance, not calcium
abundance.  Therefore, the CaT metallicities are called ${\rm
  [Fe/H]}_{\rm CaT}$.  Table~\ref{tab:metallicity} includes these
measurements alongside the synthesis measurements.  The CaT
metallicities agree within 0.5~dex of the spectral synthesis except
for star~46.  The discrepancy for that star could be due to its low
[$\alpha$/Fe] ratio.  The star could very well have an iron abundance
consistent with the synthesis measurement and a calcium abundance
consistent with the CaT measurement.  In support of this hypothesis,
the synthesis value of [Ca/H] agrees well with ${\rm [Fe/H]}_{\rm
  CaT}$.

Figure~\ref{fig:trends}a shows the mass--metallicity relation for MW
dwarf satellite galaxies \citep{kir13b,kir15b}.  The mean metallicity
of Tri~II lies above other UFDs of similar stellar mass.
Equivalently, it could be said that Tri~II is less massive than dwarf
galaxies of similar metallicity.  Interpreted in that way, it is easy
to see how Tri~II could have been tidally stripped by the MW\@.  If
the galaxy was originally more massive and it obeyed the
mass--metallicity relation, like nearly every other satellite galaxy,
then tidal stripping would remove mass but leave the mean metallicity
nearly untouched.  Thus, stripping corresponds to a leftward move in
Figure~\ref{fig:trends}a.  Segue~2 is another galaxy that appears to
have been tidally stripped \citep{kir13a}.  Like Tri~II, Segue~2 has
an unresolved velocity dispersion, and its stellar mass is lower than
most other galaxies of similar metallicity.

We computed the metallicity dispersion in the same manner as the
velocity dispersion.  We computed the likelihood \citep[Equation~2
  of][]{kir15b} that a mean metallicity and metallicity dispersion
were consistent with our six measurements of [Fe/H] and their errors.
We used an MCMC with $10^7$ links to optimize the likelihood and hence
to estimate $\sigma({\rm [Fe/H]}) =
\fehsigma^{+\fehsigmaerrupper}_{-\fehsigmaerrlower}$.  The bottom
panel of Figure~\ref{fig:sigmav} shows the probability distribution,
which is a histogram of the accepted MCMC trials.

The metallicity dispersion is largely driven by star~31 and star~46,
the binary.  If both are excluded, then the dispersion is only
marginally resolved, as shown by the cyan probability distribution in
Figure~\ref{fig:sigmav}.  An unresolved metallicity dispersion
combined with an unresolved velocity dispersion would jeopardize
Tri~II's classification as a galaxy because a dispersion in at least
one of those quantities is required to show that the stellar system
has dark matter \citep{wil12}.

Star~31 shows no apparent reason to suspect a spurious metallicity
measurement.  Star~31's metallicity measured from the from the coadded
spectrum, ${\rm [Fe/H]} = \fehmr \pm \feherrmr$, agrees with the
[Fe/H] measurements from the individual spectra: $-1.29 \pm 0.20$,
$-1.41 \pm 0.15$, and $-1.42 \pm 0.16$.  Thus, there is no apparent
problem in any individual spectrum that might lead to an anomalous
measurement of metallicity.  The only questionable aspect of star~31
is its membership (due to its distance from the center of the galaxy,
its high metallicity, and its $1.6\sigma$ deviation from the mean
velocity).

Star~46 deserves some scrutiny for its unusually low $[\alpha/{\rm
    Fe}] = \alphafebin \pm \alphafeerrbin$.  This value, measured from
the coadded spectrum, agrees with the weighted mean of the individual
spectra: $[\alpha/{\rm Fe}] = \alphafewbin \pm \alphafewerrbin$.  The
metallicity from the coadded spectrum, ${\rm [Fe/H]} = \fehbin \pm
\feherrbin$, also agrees with the weighted mean: ${\rm [Fe/H]} =
\fehwbin \pm \fehwerrbin$.  The spectrum could be contaminated by the
binary companion, but if the companion truly is an M dwarf
(Section~\ref{sec:velvar}), then it is unlikely that the faint
companion would affect the spectrum of the binary system in any
measurable way.

V17 measured chemical abundances of star~46 from a low-S/N,
high-resolution spectrum.  They measured ${\rm [Fe/H]} = -2.5 \pm
0.2$, which is inconsistent with our measurement of ${\rm [Fe/H]} =
\fehbin \pm \feherrbin$.  One possible reason for the discrepancy is
that we used ATLAS9 model atmospheres computed with an [$\alpha$/Fe]
ratio consistent with the value measured from the spectrum
\citep{kir10,kir11d}, whereas V17 used spherical MARCS model
atmospheres \citep{gus08} with solar-scaled abundances.  They also
measured low [$\alpha$/Fe] (${\rm [Mg/Fe]} = -0.7 \pm 0.3$ and ${\rm
  [Ca/Fe]} = -0.2 \pm 0.3$).  If confirmed, these abundances would
establish a dispersion in chemical abundance in Tri~II, but it would
not necessarily indicate self-enrichment by supernovae.  Variations in
light element abundances, especially Mg, are well known in globular
clusters \citep[e.g.,][]{gra04}.  Hence, if our measurement of [Fe/H]
for star~46 is erroneous \emph{and} star~31 is a non-member, then
Tri~II could very well be a globular cluster.

\begin{figure}[t!]
\centering
\includegraphics[width=\columnwidth]{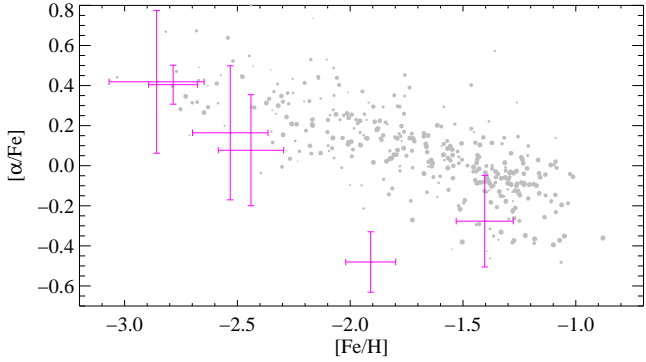}
\caption{The trend of [$\alpha$/Fe] with metallicity in Tri~II
  (\textit{magenta}) compared to the larger dwarf galaxy Sculptor
  \protect \citep[\textit{gray};][]{kir09}.  The point sizes for
  Sculptor are inversely proportional to measurement
  uncertainty.\label{fig:alpha}}
\end{figure}

Figure~\ref{fig:alpha} shows the evolution of [$\alpha$/Fe] with
increasing [Fe/H].  The decrease of [$\alpha$/Fe] is consistent with
chemical evolution dominated by Type~Ia supernovae at late times.
Type~Ia supernovae produce a great deal of iron and very little
$\alpha$ elements, thereby depressing [$\alpha$/Fe] as [Fe/H]
increases.  The decline in [$\alpha$/Fe] in Tri~II is even more severe
than in Sculptor, a larger dSph (gray points in
Figure~\ref{fig:alpha}).  The low [$\alpha$/Fe] ratios of stars~31 and
46 indicate that the galaxy's late-time chemical evolution was
dominated by Type~Ia supernovae.  Predictions of the nucleosynthetic
yield of Type~II supernovae \citep[e.g.,][]{woo95,nom06} have
significantly larger [$\alpha$/Fe].  The steep decline of
[$\alpha$/Fe] in Tri~II is similar to other UFDs \citep{var13}.  The
low [$\alpha$/Fe] ratio of star~46 is only slightly lower than stars
of similar metallicity in the Hercules UFD\@.  However, it is worth
repeating that these conclusions hinge on either the accuracy of our
measurement of [Fe/H] for star~46 or the membership of star~31.

%%%%%%%%%%%%%%%%%%%%%%%%%%%%%%%%%
%%%%%%%%%   SECTION 5   %%%%%%%%%
%%%%%%%%%%%%%%%%%%%%%%%%%%%%%%%%%

\section{Detailed Abundances of One Star}
\label{sec:detailed}

\begin{deluxetable}{lrrrrr}
\tablecolumns{6}
\tablewidth{0pt}
\tablecaption{Line List with Equivalent Widths\label{tab:ew}}
\tablehead{\colhead{Species} & \colhead{Wavelength} & \colhead{EP} & \colhead{$\log gf$} & \colhead{EW} & \colhead{$\epsilon$} \\
\colhead{} & \colhead{(\AA)} & \colhead{(eV)} & \colhead{} & \colhead{(m\AA)} & \colhead{}}
\startdata
\ion{Li}{1} & 6707.760 & 0.000 & $-0.002$ & $<  7.6$ & $< 0.65$ \\
\ion{O}{1}  & 6300.304 & 0.000 & $-9.780$ & $<  4.2$ & $< 6.77$ \\
\ion{Na}{1} & 5889.950 & 0.000 & $+0.108$ & $  96.0$ & $  2.40$ \\
\ion{Na}{1} & 5895.924 & 0.000 & $-0.194$ & $  89.5$ & $  2.62$ \\
\ion{Mg}{1} & 4057.505 & 4.350 & $-0.900$ & $  21.8$ & $  4.96$ \\
\ion{Mg}{1} & 4702.991 & 4.350 & $-0.440$ & $  46.1$ & $  4.88$ \\
\ion{Mg}{1} & 5172.684 & 2.710 & $-0.393$ & $ 193.3$ & $  4.97$ \\
\ion{Mg}{1} & 5183.604 & 2.720 & $-0.167$ & $ 204.3$ & $  4.87$ \\
\ion{Mg}{1} & 5528.405 & 4.350 & $-0.498$ & $  48.9$ & $  4.93$ \\
\ion{Al}{1} & 3961.520 & 0.010 & $-0.340$ & $ 135.1$ & $  3.53$ \\
\nodata & \nodata & \nodata & \nodata & \nodata &  \nodata \\
\enddata
\tablecomments{Wavelengths are in air.  (This table is available in its entirety in a machine-readable form in the online journal.  A portion is shown here for guidance regarding its form and content.)}
\end{deluxetable}

\begin{figure}[t!]
\centering
\includegraphics[width=\columnwidth]{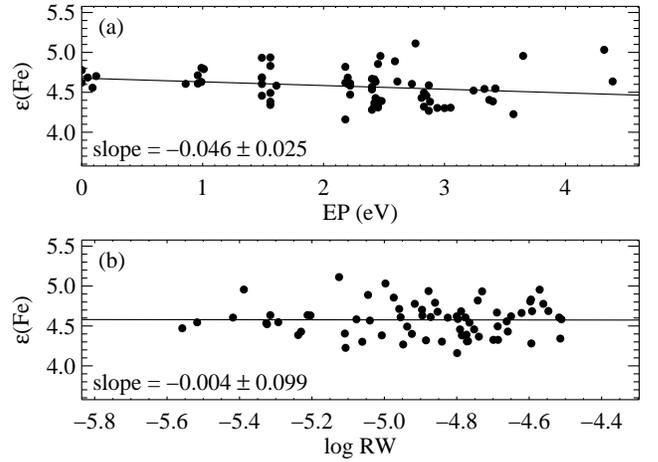}
\caption{The abundances derived from \ion{Fe}{1} lines versus
  ({\textit a}) their excitation potentials and ({\textit b}) the
  logarithms of their reduced widths (${\rm EW}/\lambda$).  The solid
  lines and the figure annotations show the best-fit linear slopes.
  The excitation balance is not perfect because we applied an
  empirical correction to the spectroscopic temperature \protect
  \citep{fre13}.\label{fig:balance}}
\end{figure}

\begin{deluxetable}{lrrrrr}
\tablecolumns{6}
\tablewidth{0pt}
\tablecaption{Abundances in TriII~40\label{tab:abund}}
\tablehead{\colhead{Element} & \colhead{$N$} & \colhead{$\epsilon$} & \colhead{[X/H]} & \colhead{[X/Fe]} & \colhead{$\sigma$}}
\startdata
\ion{Li}{1} &                      1 & $< 0.65$ & $<-0.40$ & $<+2.53$ & \nodata \\
\ion{C}{1}  & synth\tablenotemark{a} &  $ 5.42$ &  $-3.01$ &  $-0.09$ &  $0.15$ \\
\ion{O}{1}  &                      1 & $< 6.77$ & $<-1.92$ & $<+1.00$ & \nodata \\
\ion{Na}{1}\tablenotemark{b} &                      2 &  $ 2.51$ &  $-3.73$ &  $-0.81$ &  $0.15$ \\
\ion{Mg}{1} &                      5 &  $ 4.92$ &  $-2.68$ &  $+0.24$ &  $0.05$ \\
\ion{Al}{1}\tablenotemark{b} &                      1 &  $ 3.53$ &  $-2.92$ &  \phs $0.00$ & \nodata \\
\ion{Si}{1} &                      1 & $< 5.35$ & $<-2.16$ & $<+0.76$ & \nodata \\
\ion{K}{1} \tablenotemark{b} &                      1 &  $ 2.98$ &  $-2.05$ &  $+0.88$ & \nodata \\
\ion{Ca}{1} &                     16 &  $ 3.82$ &  $-2.52$ &  $+0.40$ &  $0.20$ \\
\ion{Sc}{2} &                      5 &  $ 0.53$ &  $-2.62$ &  $+0.30$ &  $0.09$ \\
\ion{Ti}{1} &                      8 &  $ 2.20$ &  $-2.75$ &  $+0.17$ &  $0.14$ \\
\ion{Ti}{2} &                     16 &  $ 2.17$ &  $-2.78$ &  $+0.14$ &  $0.16$ \\
\ion{V}{1}  &                      1 &  $ 1.03$ &  $-2.90$ &  $+0.02$ & \nodata \\
\ion{Cr}{1} &                      7 &  $ 2.27$ &  $-3.37$ &  $-0.45$ &  $0.30$ \\
\ion{Mn}{1} &                      4 &  $ 2.07$ &  $-3.36$ &  $-0.44$ &  $0.25$ \\
\ion{Fe}{1} &                     70 &  $ 4.58$ &  $-2.92$ &  \nodata &  $0.21$ \\
\ion{Fe}{2} &                     13 &  $ 4.58$ &  $-2.92$ &  \nodata &  $0.21$ \\
\ion{Co}{1} &                      1 &  $ 2.19$ &  $-2.79$ &  $+0.13$ & \nodata \\
\ion{Ni}{1} &                      4 &  $ 3.82$ &  $-2.40$ &  $+0.53$ &  $0.30$ \\
\ion{Zn}{1} &                      1 &  $ 2.04$ &  $-2.52$ &  $+0.40$ & \nodata \\
\ion{Sr}{2} &                      2 &  $-1.56$ &  $-4.43$ &  $-1.51$ &  $0.32$ \\
\ion{Y}{2}  &                      1 & $<-1.47$ & $<-3.68$ & $<-0.76$ & \nodata \\
\ion{Ba}{2} &                      1 &  $-3.11$ &  $-5.29$ &  $-2.37$ & \nodata \\
\ion{La}{2} &                      1 & $<-1.79$ & $<-2.89$ & $<+0.03$ & \nodata \\
\ion{Ce}{2} &                      1 & $<-1.12$ & $<-2.70$ & $<+0.22$ & \nodata \\
\ion{Nd}{2} &                      1 & $<-1.25$ & $<-2.67$ & $<+0.25$ & \nodata \\
\ion{Eu}{2} &                      1 & $<-2.38$ & $<-2.90$ & $<+0.02$ & \nodata \\
\ion{Dy}{2} &                      1 & $< 0.17$ & $<-0.93$ & $<+1.99$ & \nodata \\
\enddata
\tablenotetext{a}{Measured by spectral synthesis.}
\tablenotetext{b}{NLTE corrections applied to these species.}
\end{deluxetable}

We measured detailed abundances from our HIRES spectrum of star~40
using standard high-resolution abundance analysis techniques.  First,
we adopted the atomic line list of \citet{kir12c}.  The list is
described in detail by \citet{coh03}.  Second, we fit Voigt profiles
to all of the absorption lines in the list that were identifiable.  We
used local measurements of the spectral continuum within 10 line
widths of the line centers.  We fit $2\sigma$ upper limits for
undetected lines.  We found the upper limit first by calculating
$\chi_{\rm flat}^2$ for a flat spectrum.  The limit was established as
the Voigt profile with $\chi^2 = \chi_{\rm flat}^2 + 4$.
Table~\ref{tab:ew} gives the equivalent widths (EWs) of the Voigt
profiles along with the abundance determined for each line.  For
species without line detections, we present the upper limit only on
the line that yields the most stringent abundance limit.

Next, we used MOOG to compute abundances with an ATLAS9 model
atmosphere \citep{kur93,cas04}.  \citet{kir11d} constructed a grid of
ATLAS9 model atmospheres with a fine spacing in [$\alpha$/Fe].  We
used a model that was interpolated from within that grid.  We found
$T_{\rm eff}$, surface gravity, and microturbulence by minimizing the
slope of abundance versus excitation potential, minimizing the
difference in abundance between \ion{Fe}{1} and \ion{Fe}{2}, and
minimizing the slope of abundance versus reduced width (${\rm
  EW}/\lambda$).  This method is known to produce temperatures and
surface gravities that are too low \citep[e.g.,][]{the99}.  Hence, we
used the empirical temperature correction derived by \citet{fre13}.
This correction raises the temperature computed by the above method.
We then remeasured surface gravity and microturbulence by minimizing
$|\epsilon($\ion{Fe}{1}$)-\epsilon($\ion{Fe}{2}$)|$ and the slope of
abundance versus reduced width.  Figure~\ref{fig:balance} shows the
excitation potential and reduced width trends.  The latter is
consistent with zero.  The difference in iron abundance from
\ion{Fe}{1} and \ion{Fe}{2} lines is also zero.  The abundance does
depend on excitation potential because we applied the \citet{fre13}
correction to the temperature.  Table~\ref{tab:abund} gives the
abundances from the final iteration of MOOG, referenced to the solar
abundances of \citet{asp09}.

The stellar parameters measured from HIRES are $T_{\rm eff} = 4816$~K,
$\log g = 1.64$, microturbulence $\xi = 2.51$~km~s$^{-1}$, ${\rm
  [Fe/H]} = -2.92$, and $[\alpha/{\rm Fe}] = +0.25$.  The temperature
we measured from the DEIMOS spectrum is 101~K higher, and the surface
gravity is 0.25~dex higher (see Table~\ref{tab:metallicity}).  The
higher temperature led us to measure a metallicity 0.14~dex higher in
the DEIMOS spectrum compared to the HIRES spectrum.  The individual
[$\alpha$/Fe] ratios are consistent to within the errors.

The abundances of Na, Al, and K have been modified to account for
non-local thermodynamic equilibrium (NLTE) corrections.  The
corrections for the two Na~D lines are $-0.05$~dex \citep{lin11_na}.
For \ion{Al}{1}~$\lambda 3962$, we interpolated \citeauthor{and08}'s
(\citeyear{and08}) corrections in $T_{\rm eff}$ to find $+0.51$~dex.
We estimated the NLTE correction for \ion{K}{1}~$\lambda 7699$ as
$-0.21$~dex based on stars of very similar stellar parameters from
\citet{and10}.  However, \citeauthor{and10}\ caution that the NLTE
correction for potassium should be computed individually for each
star's stellar parameters and potassium EW\@.  For example, an
extrapolation in metallicity of \citeauthor{iva00}'s
(\citeyear{iva00}) tabular NLTE corrections suggests that the
correction should be $-0.49$~dex.  Weaker lines will be less affected
by NLTE corrections.  The \ion{K}{1}~$\lambda 7699$ in star~40 is
rather strong for its temperature and metallicity.  Thus, the true
correction could be stronger than what we have applied.

\begin{figure}[t!]
\centering
\includegraphics[width=\columnwidth]{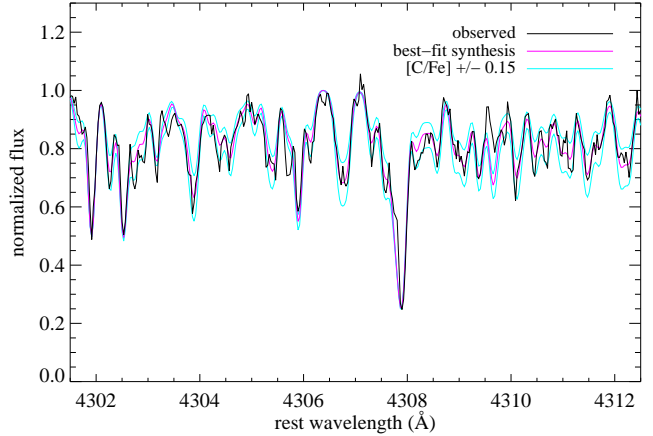}
\caption{A small region of the Keck/HIRES spectrum showing the main
  part of the G band of the CH molecule.  The magenta line shows the
  best-fit synthesis, and the cyan lines show the synthesis
  corresponding to changing the carbon abundance by our quoted error
  of $\pm 0.15$~dex.\label{fig:ch}}
\end{figure}

We measured carbon by synthesis of the Fraunhofer G band due to the CH
molecule.  We adopted the line list of \citet{jor96}.  We fit the
spectrum in the region of the G band by varying the carbon abundance
to minimize $\chi^2$.  Then, following \citet{kir12c}, we refined the
continuum determination by computing the residual between the observed
spectrum and the synthesis, fitting a polynomial to it, and then
dividing the observed spectrum by that polynomial.  We repeated the
$\chi^2$ minimization and continuum refinement until the carbon
abundance did not change between successive iterations.
Figure~\ref{fig:ch} shows the best fitting spectrum along with spectra
with the carbon abundance altered by $\pm 0.15$~dex, which is the
measurement uncertainty that we quote.  Table~\ref{tab:abund} gives
the carbon abundance that we measured.

Our high-resolution spectroscopic results for star~40 agree well with
those of V17.  We determined $T_{\rm eff}$ to be 16~K higher, $\log g$
to be 0.16~dex lower, and $\xi$ to be 0.19~km~s$^{-1}$ lower than V17.
V17 determined $T_{\rm eff}$ with the infrared flux method, whereas we
used excitation balance coupled with \citeauthor{fre13}'s
(\citeyear{fre13}) empirical temperature correction.  The close
agreement of our results supports the validity of the empirical
correction.  The metallicity agrees to within 0.05~dex.  Most other
abundances agree within the error bars with two exceptions.  First,
V17 measured an upper limit for carbon as $\epsilon({\rm C}) < 4.7$,
whereas we measured $\epsilon({\rm C}) = 5.42 \pm 0.15$.  The
difference arises from the spectral features used to measure carbon.
We used the G band at 4300~\AA, but V17 did not have access to that
region of the spectrum.  Hence, they used weaker CH features.
Figure~\ref{fig:ch} shows that our detection of carbon is clear.
Second, V17 measured $\epsilon({\rm Ti}) = 2.67 \pm 0.17$, whereas we
measured $\epsilon({\rm Ti}) = 2.18 \pm 0.16$.  Our line list for Ti
does not overlap with V17 at all.  We do not have an explanation for
the discrepancy, but we note that our Ti abundances from 24
independent lines have an interquartile range of 0.19~dex and a
standard deviation of 0.16~dex.  Also, the difference between the
average abundance from \ion{Ti}{1} and \ion{Ti}{2} lines is only
0.03~dex.

Star~40 exhibits the hallmark of detailed abundances in a UFD: an
extreme deficiency of neutron-capture elements.  Our measurement of
${\rm [Ba/Fe]} = -2.4$ is lower than found in globular clusters
\citep[e.g.,][]{iva99,sne00a,sne00b,coh11b} or in the MW halo
\citep{gra94,ful00,coh04,coh08}.  Our measurement of ${\rm [Sr/Fe]} =
-1.5$ is also very low.  High-resolution spectroscopy of most UFD
stars shows severe deficiencies in Ba and other neutron-capture
elements \citep[e.g.,][]{fre10a,fre10b,fre14,koc13,ji16d,ji16a}.  One
major exception is the UFD Reticulum~II, in which most stars have
highly super-solar [$n$/Fe] ratios \citep{ji16b,ji16c}.  One
explanation for this neutron-capture enhancement is that Reticulum~II
experienced a rare event, such as a merger between two neutron stars
or a neutron star and a black hole, that produced a great deal of
$r$-process material \citep{fre99,nis16}.  Star~40 in Tri~II has not
been affected by such an event.  Instead, Tri~II seems to have lacked
a strong source of the $r$-process, similar to most other UFDs.

The abundances of most other elements in star~40 are also typical for
dwarf galaxies.  The star is enhanced in $\alpha$ elements, consistent
with enrichment by one or more core collapse supernovae.  Lithium is
not detected.  Lithium deficiency is expected in giants because they
have undergone the first dredge-up, which mixes lithium-depleted
material from the interior of the star to the surface
\citep[e.g.][]{pil93,rya95,kir16}.  The carbon abundance is sub-solar,
which is also expected for a red giant above the luminosity function
bump in the RGB\@.  Our measurement of [C/Fe] in star~40 is consistent
with [C/Fe] for red giants of similar luminosity in classical dSphs
\citep{kir15a}.

V17 pointed out that the potassium abundance seems to be enhanced in
star~40.  This is especially noteworthy because some globular clusters
exhibit an anti-correlation between Mg and K abundances.  The
K-enhanced population in NGC~2419 shows ${\rm [K/Fe]} > +1$
\citep{coh11a,coh12,muc12}.  NGC~2808 contains a sub-population with a
much milder K enrichment \citep[${\rm [K/Fe]} \sim +0.3$,][]{muc15},
slightly larger than the average value for metal-poor halo stars
\citep[e.g.,][]{cay04}.  Prior to the discovery of K enhancement in
NGC~2419, it was thought that K was constant within every globular
cluster and at most mildly enhanced compared to the Sun \citep{tak09}.
We refer the reader to V17 for further discussion of K in star~40, but
we note that the finding of a large potassium abundance depends on the
NLTE correction.  V17 applied the same correction as we did.  Ideally,
the correction would be computed by a full NLTE calculation for this
star, as pointed out by \citet{and10}.

%%%%%%%%%%%%%%%%%%%%%%%%%%%%%%%%%
%%%%%%%%%   SECTION 6   %%%%%%%%%
%%%%%%%%%%%%%%%%%%%%%%%%%%%%%%%%%

\section{Discussion}
\label{sec:discussion}

\citet{wil12} proposed that a galaxy be defined as a stellar system
that contains dark matter.  One diagnostic of this criterion is the
observation of a velocity dispersion or rotation velocity larger than
can be explained by the baryonic mass of the galaxy.  Alternatively, a
galaxy can be identified by circumstantial evidence for dark matter,
such as the ability to retain supernova ejecta despite a stellar mass
that is too small to prevent the complete escape of such ejecta.
Stellar systems that do not pass these tests should instead be
described as star clusters.  K15a and M16 identified Tri~II as a
galaxy because they both measured a velocity dispersion significantly
in excess of what the stars alone would permit.

We have obtained new medium- and high-resolution spectroscopy of stars
in Tri~II\@.  We updated our estimate of the stellar velocity
dispersion by identifying a binary star among the original samples and
by revising M16's measurements of individual stellar velocities.  The
velocities of all but one of the \n\ member stars are consistent with
a single radial velocity within their $1\sigma$ error bars, and the
other star is consistent at $1.6\sigma$.  In other words, we cannot
resolve the velocity dispersion of Tri~II\@.  Instead, we give upper
limits: $\sigma_v < \sigmavlimone$~km~s$^{-1}$ (90\% C.L.) and
$\sigma_v < \sigmavlimtwo$~km~s$^{-1}$ (95\% C.L.).  This revision to
the velocity dispersion removes the most direct evidence for dark
matter in Tri~II\@.

The bulk properties of Tri~II's stellar population are not
particularly helpful in identifying the system as a star cluster or a
galaxy.  The half-light radius and absolute luminosity are more
consistent with galaxies than star clusters, but the two populations
overlap significantly in this space at the low luminosities of system
like Tri~II \citep{lae15a,drl15b}.  Figure~\ref{fig:trends} shows
Tri~II's properties derived from the medium-resolution spectroscopy
presented in this article.  The galaxy lies above the
mass--metallicity relation defined by Local Group dwarf galaxies.
Both globular clusters and tidally stripped dwarf galaxies occupy this
region of the mass--metallicity diagram.  The upper limits on the
kinematic properties of Tri~II are consistent with dwarf galaxies, but
they are also consistent with globular clusters, which have low
mass-to-light ratios, like stellar populations free of dark matter.

We measured individual metallicities and [$\alpha$/Fe] ratios for six
stars from medium-resolution spectroscopy.  We detected a metallicity
dispersion at high significance.  However, this conclusion depends on
one star, which happens to be the star in a binary system, or another
star, which is arguably a non-member for its large metallicity and
large radial displacement from the galaxy center.  The detailed
abundances of the binary star are unusual.  When we fit all available
Mg, Si, Ca, and Ti absorption lines in the DEIMOS spectrum
simultaneously, we measured $[\alpha/{\rm Fe}] = \alphafebin \pm
\alphafeerrbin$, which is low for a star with ${\rm [Fe/H]} = \fehbin
\pm \feherrbin$.  However, our measurement of the Mg abundance by
itself is somewhat larger: ${\rm [Mg/Fe]} = \mgfebin \pm \mgfeerrbin$.
Frustratingly, V17 measured significantly different abundances of this
star from a high-resolution but low-S/N spectrum: ${\rm [Fe/H]} = -2.5
\pm 0.2$ and ${\rm [Mg/Fe]} = -0.7 \pm 0.3$.  The conflicting results
do not lend confidence to the abundance measurements of this star.
However, the detection of a metallicity dispersion is secure
\emph{either} if star~46 has ${\rm [Fe/H]} = -1.9$ \emph{or} if
star~31 is a member.

Our high-resolution spectrum of the brightest known member star in
Tri~II shows exceptionally low abundances of Sr and Ba.  Deficiencies
in neutron-capture elements are the hallmark signature of detailed
abundances in UFDs.  Star clusters show [Sr/Fe] and [Ba/Fe] ratios
close to solar, and MW halo stars with $[n/{\rm Fe}] \lesssim -2$ are
extremely rare.  The abundances in star~40 in Tri~II are ${\rm
  [Sr/Fe]} = -1.5$ and ${\rm [Ba/Fe]} = -2.4$.  Thus, the deficiency
in neutron-capture elements in this single star may very well be the
strongest evidence in favor of classifying Tri~II as a UFD rather than
a star cluster.

\acknowledgments We thank D.\ Stern, S.\ Hemmati, and D.\ Masters for
observing the TriIIc slitmask.

We are grateful to the many people who have worked to make the Keck
Telescope and its instruments a reality and to operate and maintain
the Keck Observatory.  The authors wish to extend special thanks to
those of Hawaiian ancestry on whose sacred mountain we are privileged
to be guests.  Without their generous hospitality, none of the
observations presented herein would have been possible.

{\it Facility:} \facility{Keck:II (DEIMOS)}

\bibliography{TriII2}
\bibliographystyle{apj}

\newpage
\begin{turnpage}
\setcounter{table}{1}

\begin{deluxetable*}{llccccccccrr@{ }c@{ }lcc}
\tablewidth{0pt}
\tablecolumns{16}
\tablecaption{Radial Velocities\label{tab:velocity}}
\tablehead{\colhead{ID (K15a)} & \colhead{ID (M16)} & \colhead{RA (J2000)} & \colhead{Dec (J2000)} & \colhead{Radius} & \colhead{$(g_{\rm P1})_0$} & \colhead{$\delta g_{\rm P1}$} & \colhead{$(i_{\rm P1})_0$} & \colhead{$\delta i_{\rm P1}$} & \colhead{Masks} & \colhead{S/N\tablenotemark{a}} & \multicolumn{3}{c}{$v_{\rm helio}$} & \colhead{$\sigma(v)$} & \colhead{Member?} \\
\colhead{ } & \colhead{ } & \colhead{ } & \colhead{ } & \colhead{(arcmin)} & \colhead{(mag)} & \colhead{(mag)} & \colhead{(mag)} & \colhead{(mag)} & \colhead{ } & \colhead{(\AA$^{-1}$)} & \multicolumn{3}{c}{(km~s$^{-1}$)} & \colhead{ } & \colhead{ }}
\startdata
\nodata & 22      & 02 13 12.69 & $+36$ 08 49.4 & 2.11 & $20.71$ & $0.09$ & $20.34$ & $0.09$ &  cdefg &  44.5 & $-380.6$ & $\pm$ & $ 3.0$ &   0.6   & Y \\
128     & \nodata & 02 13 14.24 & $+36$ 09 51.1 & 1.06 & $19.93$ & $0.03$ & $19.41$ & $0.03$ &   bdeg &  25.0 & $-383.3$ & $\pm$ & $ 1.8$ &   0.4   & Y \\
116     & 21      & 02 13 15.96 & $+36$ 10 15.8 & 0.53 & $20.38$ & $0.02$ & $19.92$ & $0.02$ &  bcdeg &  26.4 & $-381.4$ & $\pm$ & $ 3.2$ &   0.9   & Y \\
106     & 40      & 02 13 16.55 & $+36$ 10 45.8 & 0.19 & $17.34$ & $0.01$ & $16.58$ & $0.01$ &   bdef & 219.5 & $-381.6$ & $\pm$ & $ 1.6$ &   0.4   & Y \\
91      & 20      & 02 13 19.32 & $+36$ 11 33.3 & 0.93 & $20.33$ & $0.03$ & $19.79$ & $0.03$ &   bcfg &  29.7 & $-380.1$ & $\pm$ & $ 4.9$ &   1.7   & Y \\
76      & 23      & 02 13 20.61 & $+36$ 09 46.5 & 1.12 & $20.83$ & $0.06$ & $20.53$ & $0.06$ &    bcg &  17.0 & $-385.2$ & $\pm$ & $ 4.2$ &   1.3   & Y \\
\nodata & 27      & 02 13 21.35 & $+36$ 08 29.1 & 2.36 & $21.30$ & $0.07$ & $21.27$ & $0.07$ &     cd &  20.0 & $-376.8$ & $\pm$ & $11.7$ &   1.6   & Y \\
65      & 46      & 02 13 21.54 & $+36$ 09 57.4 & 1.11 & $19.03$ & $0.01$ & $18.42$ & $0.01$ &   bdfg &  81.8 & $-381.0$ & $\pm$ & $ 5.9$ &   3.0   & Y \\
\nodata & 24      & 02 13 22.00 & $+36$ 10 25.9 & 0.97 & $21.22$ & $0.07$ & $21.14$ & $0.07$ &      d &  17.8 & $-370.4$ & $\pm$ & $17.1$ & \nodata & Y \\
\nodata & 26      & 02 13 24.83 & $+36$ 10 21.8 & 1.54 & $21.40$ & $0.11$ & $21.17$ & $0.11$ &      c &  19.9 & $-375.6$ & $\pm$ & $11.2$ & \nodata & Y \\
\nodata & 9       & 02 13 27.33 & $+36$ 13 30.5 & 3.45 & $21.25$ & $0.10$ & $21.05$ & $0.10$ &      d &  17.6 & $-387.6$ & $\pm$ & $ 7.7$ & \nodata & Y \\
\nodata & 29      & 02 13 30.95 & $+36$ 11 56.0 & 3.00 & $21.96$ & $0.20$ & $21.68$ & $0.20$ &      c &  14.0 & $-386.2$ & $\pm$ & $ 4.7$ & \nodata & Y \\
\nodata & 31      & 02 13 52.66 & $+36$ 13 24.1 & 7.61 & $20.63$ & $0.03$ & $20.12$ & $0.03$ &    cdg &  42.8 & $-377.1$ & $\pm$ & $ 2.7$ &   0.9   & Y \\
\nodata & 25      & 02 13 17.14 & $+36$ 07 14.1 & 3.47 & $21.15$ & $0.05$ & $21.07$ & $0.05$ &     cd &  21.3 & \multicolumn{3}{c}{\nodata} & \nodata & ?\tablenotemark{b} \\
\nodata & 1       & 02 12 46.71 & $+36$ 07 58.2 & 6.77 & $19.63$ & $0.01$ & $19.21$ & $0.01$ &      e &  49.2 & $ -85.4$ & $\pm$ & $ 1.8$ & \nodata & N \\
\nodata & 2       & 02 12 50.67 & $+36$ 08 24.6 & 5.86 & $19.76$ & $0.02$ & $19.27$ & $0.02$ &      e &  55.0 & $ -89.8$ & $\pm$ & $ 2.1$ & \nodata & N \\
\nodata & 14      & 02 13 02.68 & $+36$ 10 32.9 & 2.97 & $19.19$ & $0.01$ & $18.76$ & $0.01$ &      e &  71.7 & $ -40.6$ & $\pm$ & $ 1.6$ & \nodata & N \\
\nodata & 34      & 02 13 08.87 & $+36$ 05 44.3 & 5.25 & $18.38$ & $0.01$ & $17.90$ & $0.01$ &      d & 122.5 & $ -40.7$ & $\pm$ & $ 1.5$ & \nodata & N \\
166     & \nodata & 02 13 11.42 & $+36$ 12 32.4 & 2.21 & $18.51$ & $0.04$ & $17.71$ & $0.04$ &      b & 115.8 & $  11.0$ & $\pm$ & $ 5.9$ & \nodata & N \\
174     & \nodata & 02 13 12.86 & $+36$ 11 20.1 & 1.12 & $18.22$ & $0.01$ & $17.87$ & $0.01$ &      b & 102.7 & $-100.8$ & $\pm$ & $ 1.6$ & \nodata & N \\
177     & 4       & 02 13 13.22 & $+36$ 11 35.4 & 1.23 & $19.83$ & $0.01$ & $19.52$ & $0.01$ &     be &  36.7 & $-278.0$ & $\pm$ & $ 3.6$ &   1.4   & N \\
126     & 15      & 02 13 14.11 & $+36$ 12 22.5 & 1.80 & $17.19$ & $0.00$ & $16.33$ & $0.00$ &      b & 127.2 & $   2.4$ & $\pm$ & $ 1.5$ & \nodata & N \\
127     & \nodata & 02 13 14.32 & $+36$ 13 03.5 & 2.45 & $19.63$ & $0.01$ & $19.08$ & $0.01$ &      b &  54.5 & $ -86.0$ & $\pm$ & $ 1.8$ & \nodata & N \\
113     & 19      & 02 13 16.18 & $+36$ 11 16.3 & 0.62 & $19.15$ & $0.01$ & $18.63$ & $0.01$ &   bdfg &  72.7 & $ -56.0$ & $\pm$ & $ 2.7$ &   1.4   & N \\
111     & \nodata & 02 13 16.41 & $+36$ 13 00.8 & 2.33 & $19.06$ & $0.01$ & $18.71$ & $0.01$ &     bf &  66.7 & $ -61.3$ & $\pm$ & $ 1.8$ &   0.5   & N \\
\nodata & 47      & 02 13 17.33 & $+36$ 08 18.3 & 2.40 & $18.72$ & $0.02$ & $18.32$ & $0.02$ &  cdefg &  96.7 & $ -71.6$ & $\pm$ & $ 1.6$ &   0.4   & N \\
100     & \nodata & 02 13 18.03 & $+36$ 12 32.9 & 1.85 & $18.99$ & $0.01$ & $18.23$ & $0.01$ &     bf &  87.2 & $ -36.4$ & $\pm$ & $ 2.8$ &   1.5   & N \\
84      & \nodata & 02 13 19.74 & $+36$ 11 15.3 & 0.72 & $19.41$ & $0.01$ & $18.89$ & $0.01$ &      b &  60.7 & $ -66.4$ & $\pm$ & $ 1.7$ & \nodata & N \\
82      & 44      & 02 13 19.89 & $+36$ 12 12.3 & 1.58 & $18.69$ & $0.01$ & $18.01$ & $0.01$ &     bf & 101.8 & $-176.9$ & $\pm$ & $ 1.5$ &   0.1   & N \\
45      & 42      & 02 13 24.28 & $+36$ 10 15.1 & 1.45 & $17.32$ & $0.00$ & $16.67$ & $0.00$ &      b & 153.0 & $ -62.1$ & $\pm$ & $ 1.5$ & \nodata & N \\
\nodata & 45      & 02 13 33.37 & $+36$ 10 03.1 & 3.29 & $18.86$ & $0.01$ & $18.42$ & $0.01$ &   cdeg & 113.9 & $ -59.7$ & $\pm$ & $ 2.6$ &   1.4   & N \\
\nodata & 3       & 02 13 38.73 & $+36$ 15 02.0 & 6.10 & $19.73$ & $0.02$ & $19.30$ & $0.02$ &    cdg &  56.3 & $ -67.8$ & $\pm$ & $ 1.6$ &   0.2   & N \\
\nodata & 50      & 02 13 39.17 & $+36$ 10 39.3 & 4.39 & $18.83$ & $0.00$ & $18.35$ & $0.00$ &      e &  83.5 & $ -29.8$ & $\pm$ & $ 1.6$ & \nodata & N \\
\nodata & 18      & 02 13 47.91 & $+36$ 16 15.1 & 8.28 & $18.82$ & $0.01$ & $18.18$ & $0.01$ &     dg &  95.2 & $-149.1$ & $\pm$ & $ 1.7$ &   0.6   & N \\
\enddata
\tablenotetext{a}{To convert to S/N per pixel, multiply by 0.57.}
\tablenotetext{b}{This star is a double-lined spectroscopic binary.  The velocities of the individual components straddle the mean radial velocity of Tri~II, but the two epochs were not sufficient to measure the center-of-mass velocity reliably.}
\end{deluxetable*}

\setcounter{table}{3}

\begin{deluxetable*}{llccccccccc}
\tablewidth{0pt}
\tablecolumns{11}
\tablecaption{Chemical Abundances\label{tab:metallicity}}
\tablehead{\colhead{ID (K15a)} & \colhead{ID (M16)} & \colhead{$T_{\rm eff}$} & \colhead{$\log g$} & \colhead{[Fe/H]$_{\rm CaT}$} & \colhead{[Fe/H]} & \colhead{[Mg/Fe]} & \colhead{[Si/Fe]} & \colhead{[Ca/Fe]} & \colhead{[Ti/Fe]} & \colhead{[$\alpha$/Fe]} \\
\colhead{ } & \colhead{ } & \colhead{(K)} & \colhead{(cm~s$^{-2}$)} & \colhead{ } & \colhead{ } & \colhead{ } & \colhead{ } & \colhead{ } & \colhead{ } & \colhead{ }}
\startdata
128     & \nodata & 5466 & 3.17 & $-3.01 \pm 0.08$ & $-2.44 \pm 0.15$ & $+0.46 \pm 0.83$ &     \nodata      & $+0.39 \pm 0.22$ &     \nodata      & $+0.08 \pm 0.28$ \\
116     & 21      & 5635 & 3.41 & $-3.14 \pm 0.13$ & $-2.86 \pm 0.21$ & $+1.07 \pm 0.67$ & $+0.59 \pm 0.50$ & $+0.24 \pm 0.36$ & $+1.00 \pm 0.44$ & $+0.42 \pm 0.36$ \\
106     & 40      & 4917 & 1.89 & $-2.71 \pm 0.02$ & $-2.78 \pm 0.11$ & $+0.59 \pm 0.38$ & $+0.43 \pm 0.17$ & $+0.45 \pm 0.12$ & $+0.27 \pm 0.14$ & $+0.40 \pm 0.10$ \\
91      & 20      & 5434 & 3.33 & $-2.13 \pm 0.06$ & $-2.53 \pm 0.17$ &     \nodata      & $+0.50 \pm 0.35$ & $-0.27 \pm 0.83$ & $+0.71 \pm 0.42$ & $+0.16 \pm 0.33$ \\
65      & 46      & 5282 & 2.74 & $-2.56 \pm 0.04$ & $-1.91 \pm 0.11$ & $+0.21 \pm 0.28$ &     \nodata      & $-0.39 \pm 0.15$ & $-0.79 \pm 0.76$ & $-0.48 \pm 0.15$ \\
\nodata & 31      & 5551 & 3.49 & $-0.90 \pm 0.07$ & $-1.40 \pm 0.13$ & $+0.89 \pm 0.26$ & $-0.09 \pm 0.28$ & $-0.37 \pm 0.25$ &     \nodata      & $-0.28 \pm 0.23$ \\
\enddata
\end{deluxetable*}

\end{turnpage}

\end{document}